\documentclass[a4paper,10pt,twocolumn]{article}

\usepackage[english]{babel}
\usepackage[utf8]{inputenc}
\usepackage[T1]{fontenc}

\usepackage[top=1.5cm, left=1.5cm, right=1.5cm, bottom=1.5cm]{geometry}

\renewenvironment{abstract}{\bf\small {\em\ Abstract---}}{}

\usepackage{amsfonts,amssymb,amsmath,amsthm}
\usepackage{subfigure}
\usepackage{graphicx}
\usepackage[footnotesize]{caption}

\usepackage[hidelinks]{hyperref}
\usepackage{amsfonts,amssymb,amsmath,amsthm}
\usepackage[style=numeric,sorting=none,hyperref=true,backend=biber,url=false,maxnames=10]{biblatex}
\usepackage{bm}
\usepackage{xcolor}
\usepackage{algorithm}
\usepackage{algorithmic}
\algsetup{linenodelimiter=\ }
\algsetup{linenosize=\scriptsize}
\algsetup{indent=1.5em}
\definecolor{gray}{gray}{0.50}

\title{Generalised Approximate Message Passing for Non-I.I.D. Sparse Signals}

\author{Christian Schou Oxvig$^1$ and Thomas Arildsen$^2$.\\
  \footnotesize $^1$ Formerly also at: \
  $^2$\begin{minipage}[t]{.25\linewidth}
    Aalborg University\\
    Technical Faculty of IT \& Design\\
    Department of Electronic Systems
  \end{minipage}}
\date{\empty} 

\begin{document}

\maketitle

\begin{abstract}
  Generalised approximate message passing (GAMP) is an approximate
  Bayesian estimation algorithm for signals observed through a linear
  transform with a possibly non-linear subsequent measurement
  model. By leveraging prior information about the observed signal,
  such as sparsity in a known dictionary, GAMP can for example
  reconstruct signals from under-determined measurements -- known as
  compressed sensing. In the sparse signal setting, most existing
  signal priors for GAMP assume the input signal to have
  i.i.d. entries. Here we present sparse signal priors for GAMP to
  estimate non-i.d.d. signals through a non-uniform weighting of the
  input prior, for example allowing GAMP to support model-based
  compressed sensing.
\end{abstract}

\section{Introduction}
\label{sec:introduction}

Generalised approximate message passing (GAMP) was introduced by
\citeauthor{Rangan2011} in \cite{Rangan2011,Rangan2011Full}. GAMP
addresses the estimation of signals $\mathbf x$ observed through a
linear transform as follows
\begin{equation}
  \label{eq:signal-model}
  \mathbf y = \mathbf{A x} + \mathbf{e}
\end{equation}
where
$\mathbf y \in \mathbb C^{m}, \mathbf A \in \mathbb C^{m \times n},
\mathbf x \in \mathbb C^{n}, \mathbf e \in \mathbb C^{m}$.
Here we express $\mathbf e$ as an additive noise which is classical in
linear measurement models. We define the intermediate measurements
without noise:
\begin{equation}
  \label{eq:noiseless-meas}
  \mathbf z = \mathbf{A x},\quad \mathbf y = \mathbf{z} + \mathbf{e}
\end{equation}
The noise term $\mathbf e$ does not have to be a strictly additive
term independent of $\mathbf x$. It can more generally be a (possibly
non-linear) separable measurement channel expressed through a p.d.f.\
of $\mathbf y$;
$p(\mathbf y | \mathbf{z}; \bm \theta_O) = \prod_{l=0}^{m-1} p(y_l |
[\mathbf z]_l; [\bm \theta_O]_l)$ where $[\bm \theta_O]_l$ represents the parameters of the channel.

GAMP can be used quite generally for estimation of signals from many
different distributions of $\mathbf x$. Here we consider the
compressed sensing setting (see \cite{Candes2008}) where $m < n$,
i.e. an under-determined system which may be solved if
\begin{equation}
  \label{eq:signal-sparsity}
  \|\mathbf x\|_0 = k \ll n,\quad k < m,
\end{equation}
where the $\|\mathbf x\|_0$ operator counts the number of non-zero
entries in $\mathbf x$. GAMP can solve~\eqref{eq:signal-model} in the
compressed sensing setting when a sparse prior can be imposed on
$\mathbf x$ to model~\eqref{eq:signal-sparsity}.

For the compressed sensing setting, the algorithm approximate message
passing (AMP) was proposed to estimate $\mathbf x$ with an
i.i.d. Laplacian prior and i.i.d additive white Gaussian noise
$\mathbf e$ \cite{Donoho2009}. GAMP can be seen as a generalisation of
AMP that allows for a wider range of probability distributions on the
signals $\mathbf x$ and on the measurements $\mathbf y$ given $\mathbf
x$.

Another prior that can be used to model sparse signals in Bayesian
estimators such as GAMP is the so-called spike-and-slab model
\cite{Mitchell1988}. According to this model, each entry $x_j$ in
$\mathbf x,\ j = 0, 1,\ldots n$ is distributed according to a linear
combination of a Dirac delta p.d.f $\delta(x_j)$ and another p.d.f.\
$\phi(x_j; \bm \theta_I)$:
\begin{equation}
  \label{eq:spike-and-slab}
  p(x_j; \tau, \bm \theta_I) = (1 - \tau) \delta(x_j) +
  \tau \phi(x_j; \bm \theta_I)
\end{equation}
The $\delta(x)$ models the fact that many of the entries $x_i$ are
zero (by the sparsity of $\mathbf x$). The variable $0 \le \tau \le 1$
controls the sparsity of $\mathbf x$, i.e. how likely the entries are
to be zero. The function $\phi(x_j; \bm \theta_I)$ can be chosen to
represent the p.d.f.\ of entries of $\mathbf x$ that are not zero;
$\bm \theta_I$ represents the parameters of $\phi$.

One example of such a spike-and-slab prior is the Bernoulli-Gauss
distribution~\cite{Vila2011}:
\begin{equation}
  \label{eq:bernoulli-gauss}
  p(x_j; \tau, \mu, \sigma) = (1 - \tau) \delta(x_j) +
  \tau \mathcal N(x_j; \mu, \sigma),
\end{equation}
where the p.d.f.\ representing the non-zero entries is the Gaussian
distribution with mean $\mu$ and std. deviation $\sigma$
(corresponding to the parameters $\bm \theta_I$). The term
Bernoulli-Gauss (BG) reflects the fact that each entry $x_i$ can be
seen as the product of a Bernoulli random variable (values 0 or 1) and
a Gaussian random variable.

Applying GAMP with the signal prior \eqref{eq:bernoulli-gauss} assumes
that the entries of $\mathbf x$ are i.i.d. In many cases, signals of
interest exhibit additional structure that can be exploited to
estimate them more accurately \cite{Baraniuk2010}.

Many different algorithmic approaches to modelling and leveraging such
signal structure can be taken. Schniter et al.  have for example
produced substantial results on incorporating the ability to learn the
structure of non-i.i.d. priors into the GAMP framework, see
e.g. \cite{Ziniel2012}. We take a different approach here and propose
a weighted spike-and-slab model that can be used to model
non-i.i.d. signals. We present the corresponding GAMP equations
derived for this prior model on $\mathbf x$ as well as results from numerical simulations that show the improvements in reconstruction capabilities that are achievable using such a structured prior.

\section{Weighted-Prior GAMP}
\label{sec:wgamp}

We re-state the uniform variance MMSE GAMP algorithm \cite{Rangan2011Full} in
Algorithm~\ref{alg:schniter_mmse_gamp_rangan_sum_approx} as given in
\cite{Oxvig2017} with some variable changes to
match~\eqref{eq:signal-model}.
\begin{algorithm}[htbp]
  \newcommand{\tmo}{t\mbox{-}1}
  \begin{algorithmic}[1]
    \STATE \textbf{initialise:}\\ $\bar{\mathbf{x}}_{0} = \mathbb{E}_{\mathbf{x}|\bm{\theta}_I}[\mathbf{x}]$, $\breve{x}_{0} = \frac{1}{n}\sum\left(\text{Var}_{\mathbf{x}|\bm{\theta}_I}(\mathbf{x})\right)$, $\mathbf{q}_0 = \mathbf{0}_m$
    \FOR{$t = 1 \dots T_{\text{max}}$}
    \STATE $\breve{v}_{t} = \frac{1}{m}||\mathbf{A}||_F^2\breve{x}_{\tmo}$
    \STATE $\mathbf{o}_{t} = \mathbf{A}\bar{\mathbf{x}}_{\tmo} - \breve{v}_{t}\mathbf{q}_{\tmo}$
    \STATE $\bar{\mathbf{z}}_{t} = f_{\bar{z}}(\breve{v}_{t}, \mathbf{o}_{t}; \mathbf{y}, \bm{\theta}_{o})$
    \STATE $\tilde{\mathbf{z}}_{t} = f_{\tilde{z}}(\breve{v}_{t}, \mathbf{o}_{t}; \mathbf{y}, \bm{\theta}_{o})$
    \STATE $\mathbf{q}_{t}$ = $\frac{\bar{\mathbf{z}}_{t} - \mathbf{o}_{t}}{\breve{v}_{t}}$
    \STATE $\breve{u}_{t}$ = $\frac{1}{m}\sum\left(\frac{\breve{v}_{t} - \tilde{\mathbf{z}}_{t}}{\breve{v}_{t}^2}\right)$
    \STATE $\breve{s}_{t} = [\frac{1}{n}||\mathbf{A}||_F^2\breve{u}_{t}]^{-1}$
    \STATE $\mathbf{r}_{t} = \bar{\mathbf{x}}_{\tmo} + \breve{s}_{t}\mathbf{A}^H\mathbf{q}_{t}$
    \STATE $\bar{\mathbf{x}}_{t} = f_{\bar{x}}(\breve{s}_{t}, \mathbf{r}_{t}; \bm{\theta}_{I})$
    \STATE $\breve{x}_{t} = \frac{1}{n}\sum\left(f_{\tilde{x}}(\breve{s}_{t}, \mathbf{r}_{t}; \bm{\theta}_{I})\right)$
    \IF{stop criterion is met}
    \STATE \textbf{break}
    \ENDIF
    \ENDFOR
  \end{algorithmic}
  \caption{- Uniform variance MMSE GAMP}
  \label{alg:schniter_mmse_gamp_rangan_sum_approx}
\end{algorithm}
We stress that our proposed prior may as well be used with the non-uniform variants of GAMP that track individual variances. Due to limited space, a detailed explanation of the algorithms can be
found in \cite{Oxvig2017} and references therein.

In order to apply the GAMP algorithm with a specific entry-wise input
prior $p(x_j; \tau, \bm \theta_I)$ and measurement channel function
$p(\mathbf y | \mathbf{z}; \bm \theta_O)$, it is necessary to derive
the posterior p.d.f.
$p(x_j | \mathbf y; s_j, r_j, [\bm \theta_I]_j)$. The evaluation of
input posterior and measurement channel functions is incorporated in
the GAMP iterations in the form of $\bar{\mathbf x}_t$ and
$\breve x_t$ in
Algorithm~\ref{alg:schniter_mmse_gamp_rangan_sum_approx}, lines 11-12,
respectively $\bar{\mathbf z}_t$ and $\tilde{\mathbf z}_t$ in lines
5-6. We refer to these functions, $f_{\bar{x}}, f_{\tilde{x}}$ and $f_{\bar{z}}, f_{\tilde{z}}$ as the GAMP
input- and output channnel functions, respectively.

Channel functions for the BG input channel~\eqref{eq:bernoulli-gauss}
can be found in~\cite{Krzakala2012a,JasonParker_PhDThesis,Vila2011}
and for the additive white Gaussian noise output channel
in~\cite{Rangan2011,Rangan2011Full}.

Here we propose a modified Bernoulli-Gauss input channel that supports
non-uniform sparsity over the signal $\mathbf x$, i.e. different
entries $x_j$ can have different probabilities of being zero. We
extend the spike-and slab model~\eqref{eq:spike-and-slab} as
follows~\cite[p. 15]{Oxvig2017}:
\begin{equation}
  \label{eq:weighted-spike-and-slab}
  p(x_j; \tau, \bm \theta_I) = (1 - w_j \tau) \delta(x_j) +
  w_j \tau \phi(x_j; \bm \theta_I)
\end{equation}
General input channel functions have been derived for this
model~\cite[p. 16]{Oxvig2017}. As one example of such a weighted
spike-and-slab input channel, we have derived the following
closed-form expressions for the weighted BG input channel
functions~\cite[p. 26]{Oxvig2017}, i.e. where $\phi(x_j; \bm \theta_I)
= \mathcal N(x_j; \bar\theta, \tilde\theta)$:
\begin{align}
  &f_{\bar{x}_j}(s_j,r_j;\bm{\theta}_I)
  =
    \pi_j^{\text{w}}(r_j, s_j, \bm{\theta}_I)\left(
    \frac{\frac{\bar{\theta}}{\tilde{\theta}} +
    \frac{r_j}{s_j}}{\frac{1}{\tilde{\theta}} + \frac{1}{s_j}}
    \right) \label{eq:weighted_iidsgb_input_channel_mean} \\
  &f_{\tilde{x}_j}(s_j,r_j;\bm{\theta}_I)
  =
    \pi_j^{\text{w}}(r_j, s_j, \bm{\theta}_I) \notag\\
  &\ \ \ \cdot \left(
    \frac{1}{\frac{1}{\tilde{\theta}} + \frac{1}{s_j}} +
    \left(\frac{\frac{\bar{\theta}}{\tilde{\theta}} +
    \frac{r_j}{s_j}}{\frac{1}{\tilde{\theta}} +
    \frac{1}{s_j}}\right)^2 \right)
    - f_{\bar{x}_j}(s_j,r_j;\bm{\theta}_I)^2
    \label{eq:weighted_iidsgb_input_channel_variance}
\end{align}
where the function $\pi_j^{\text{w}}$ is given in eqs.~{(3.46)-(3.50)}
in~\cite[p. 15]{Oxvig2017} and $\bar{\theta}$ and $\tilde{\theta}$
are the mean and the variance, respectively, of the Gaussian term in
each entry of the GAMP input channel.

With the proposed weighted input
prior~\eqref{eq:weighted-spike-and-slab} it is possible to model
signals with varying probability of zero entries across the signal
$\mathbf x$. Additionally, similar to the mechanism for the i.i.d. BG
input channel in~\cite{Vila2011}, we have derived formulas for
updating the parameters $\tau, \bm{\theta}_I,$ and $\bm{\theta}_O$
using expectation-maximization (EM) as part of the GAMP
algorithm~\cite[sec. 6]{Oxvig2017}. We retain separate parameters
$\tau$ and $w_j$ to enable estimating the overall sparsity via EM
without too many free parameters.

Next, we demonstrate by numerical experiments how the proposed model
can improve estimation of compressively sensed signals with a known
non-uniform sparsity structure.

\section{Numerical Example}
\label{sec:imaging-example}

We perform a set of numerical experiments to demonstrate the benefits
of the proposed weighted sparse prior for GAMP when such a model
matches the signal of interest. We simulate the algorithm's
reconstruction capabilities in the form of a phase transition diagram
over the full sparsity / under-sampling parameter
space. See~\cite{Donoho2009a} for an introduction to the phase
transition in compressed sensing problems.

We simulate sparse BG signals according to the proposed weighted
model~\eqref{eq:weighted-spike-and-slab} where the weights $w_j$ are
selected to have a Gaussian shape over the support of the signal
vector $\mathbf x$:
\begin{equation}
  \label{eq:exp-weights}
  w_j = e^{-25 \left(\frac{j}{n} - \frac12\right)^2},\quad j = 0,1,\ldots,n
\end{equation}

\begin{figure}[th]
  \centering
  \includegraphics[width=\linewidth]{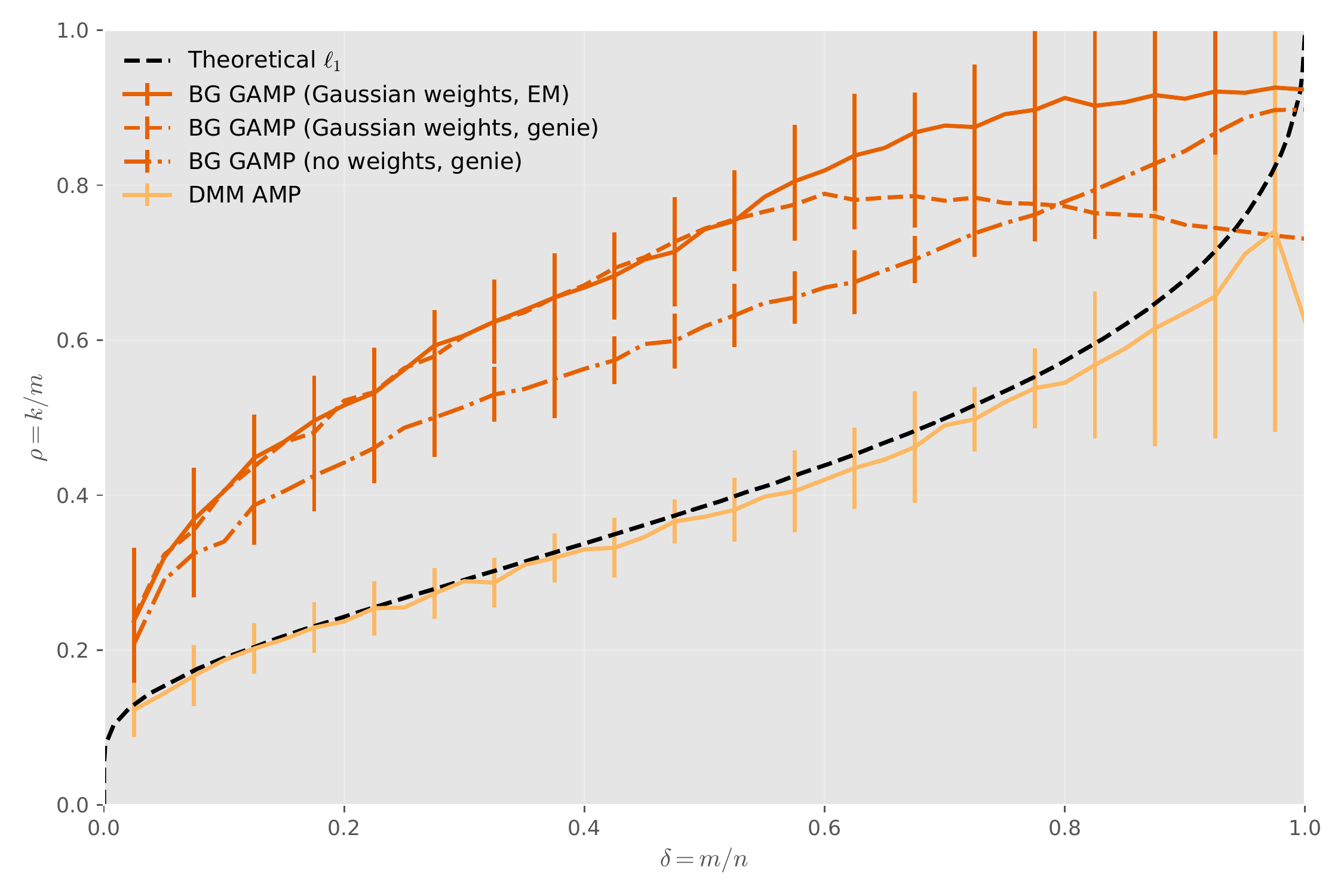}
  \caption{Numerical phase transition simulation results.}
  \label{fig:phase-trans}
\end{figure}

We partition the parameter space of $\delta = m/n$ and $\rho = k/m$
into a grid of points in each of which we reconstruct 10 random
signals $\mathbf x$ with $k$ non-zero entries on average, measured
with a randomly generated matrix $\mathbf A$ with i.i.d. random
Gaussian entries. The reconstruction success (exact to within a small
tolerance) of each signal is evaluated and the success rates in the
$(\delta, \rho)$ points are used to estimate the phase transitions
shown in Figure~\ref{fig:phase-trans}. The results shown here are a
small subset of the simulation results available
along with source code in~\cite{Oxvig2017_results}.

For reference, the figure includes the theoretical $\ell_1$-optimisation phase
transition curve~\cite{Donoho2009a}, simulation results for a non-weighted BG prior~\eqref{eq:bernoulli-gauss} as well as simulation results for the
implied Laplacian model used by AMP, \textit{``DMM AMP''}~\cite{Donoho2009}. We present results for a weighted prior that matches our Gaussian weights in~\eqref{eq:exp-weights} for: 1. The algorithm knows the true model parameters (\textit{genie}) and 2. EM is used with a re-weighting scheme to estimate the parameters~\cite[sec. 6]{Oxvig2017}.  Comparing the two, we can see that knowing the underlying sparsity structure of the signal of interest improves the reconstruction capabilities substantially compared to simply assuming no weighting. Note, however, that it is important to allow the GAMP algorithm some ``slack'' in the form of estimating the parameters $\tau, \bar{\theta}_j$, and $\tilde{\theta}_j$ using EM to get the best performance. The error bars shown on the result curves correspond to the 10\% to 90\% percentile range of the logistic sigmoid functions fitted to the reconstruction outcomes to estimate the phase transition.

\section{Conclusion}
\label{sec:conclusion}

We have proposed a model for a class of non-uniformly structured
sparse signals for use in the generalised approximate message passing
(GAMP) algorithm. The proposed approach models sparse signals where
the probability of zero vs. non-zero entries can vary across the
support of the signal of interest. We have demonstrated through
numerical examples how exploiting such structure when present in
signals can substantially improve reconstruction of compressively
sensed signals.

\printbibliography

\end{document}